\documentclass[12pt,preprint]{aastex}
\begin{document}

\title{Boron Depletion in F and G Dwarf Stars \\
and the Beryllium-Boron Correlation\altaffilmark{1} }

\author{Ann Merchant Boesgaard\altaffilmark{2}} \affil{Institute for
Astronomy, University of Hawai`i at M\-anoa, \\ 2680 Woodlawn Drive, Honolulu,
HI {\ \ }96822 \\ }
\email{boes@ifa.hawaii.edu}

\author{Constantine P. Deliyannis}
\affil{Department of Astronomy, Indiana University\\
727 East 3rd Street, Swain Hall West 319, Bloomington, IN {\ \ }47405-7105}
\email{con@athena.astro.indiana.edu, }

\and

\author{Aaron Steinhauer}
\affil{Department of Astronomy, University of Florida\\
211 Bryant Space Science Center, Gainesville, FL {\ \ }32611-2055}
\email{aarons@astro.ufl.edu}

\altaffiltext{1}{Based on observations obtained with the NASA/ESA {\it Hubble
Space Telescope} through the Space Telescope Science Institute which is
operated by the Association of Universities for Research in Astronomy, Inc.,
under NASA contract NAS5--26555.}

\altaffiltext{2}{Visiting Astronomer, W.M. Keck Observatory, jointly operated
by the California Institute of Technology and the University of California}

\begin{abstract}
Boron provides a special probe below the stellar surface as it survives to
greater depths than do Li and Be.  To search for B depletions we have observed
B in 13 F and G dwarfs with large Be depletions; for comparison we also
obtained spectra of five stars which are undepleted in Li and Be.  We have
used HST with STIS to obtain spectra of the B I resonance line at 2497 \AA.
The spectral resolution is 30,000 or 114,000 and the median signal-to-noise
ratio is 70 per pixel.  New Be and Li spectra have been obtained at Keck I
with HIRES of four of the five standard stars at $\sim$48,000 resolution.
Abundances have been determined by the spectrum synthesis method with MOOG.  A
comparison between the standard stars and those with severe Be depletions
shows a distinct difference in the B abundances between the two groups of 0.22
dex.  We have discovered a correlation between the Be and B abundances.  The
slope between A(Be) and A(B)$_{NLTE}$ is 0.22 $\pm$0.05 (where A(element) =
log N(element)/N(H) + 12.00) which, as expected, is shallower than the slope
between A(Li) and A(Be) of 0.38.  We have normalized the light element
abundances to account for the observation that the initial abundances are
somewhat lower in lower metallicity stars by employing recently published
empirical relations between Be and [Fe/H] and between B and [Fe/H] (Boesgaard
et al).  The correlation between the normalized A(Be) and A(B)$_{NLTE}$ has a
slope of 0.18 $\pm$0.06.  The star with the largest Be depletion, HR 107 a
main sequence Ba star, also has the largest B depletion with the B abundance
lower by a factor of 3.5 relative to the standard stars.
\end{abstract}


\section{Introduction}

Each of the three light elements, lithium (Li), beryllium (Be), and boron (B),
survives to different depths in models of cool stars.  Boron is destroyed at
temperatures higher than 5 $\times$ 10$^6$ K and so provides a probe into
deeper layers than do Li (destroyed at T $>$ 2.5 $\times$ 10$^6$ K) and Be
(destroyed at T $>$ 3.5 $\times$ 10$^6$ K).  The amount of each element on the
surface of a star is decreased if the stellar surface material is circulated
down inside the star to its individual critical temperature.  Although
standard stellar models of F and G dwarfs indicate there would be no such
circulation and thus no depletion of Li, Be, and B, observations show
substantial element deficiencies.  Several mechanisms that would give rise to
additional mixing below the surface convection zones of such stars have
been proposed to explain the observed depletions.

The most dramatic observational conflict with the model predictions is the
pronounced Li deficiencies in the narrow region where stellar effective
temperatures ($T_{\rm eff}$) are between 6300 and 6850 K.  This ``Li dip'' is
especially well-delineated in the Hyades star cluster, first found by
Boesgaard \& Tripicco (1986).  A similar ``Be dip'' has recently been found
for the Hyades by Boesgaard \& King (2002) (see also Boesgaard, Armengaud \&
King 2004a).  Several studies have been made of this phenomenon in other star
clusters.  One key result is that there is little or no Li-dip in the young
(70 Myr) Pleiades cluster (Pilachowski, Booth \& Hobbs 1987, Boesgaard, Budge
\& Ramsay 1988), and no Be-dip (Boesgaard, Armengaud \& King 2003a).  Thus we
understand that the depletions of Li and Be in the dip stars occurs during
main-sequence evolution, not during the pre-main sequence phase.  The effect
of the additional depletion shows up in Li and Be deficiencies in
the surface abundances by stellar ages of a few hundred million years.

For both cluster stars and field stars we see that the depletions of Li and Be
are correlated in the temperature regime of 5900 - 6650 K, corresponding
roughly to the bottom of the Li-Be dip ($\sim$6650 K) to the Li-Be peak
($\sim$5900 K) in the early G dwarfs (see e.g. Boesgaard et al.~2004c).  The
slope of the linear relation between A(Li) and A(Be) is elegantly matched to
the predictions of rotationally-induced mixing models of Deliyannis \&
Pinsonneault (1997) as can been seen in Figure 12 of Boesgaard et al.~(2004c).
(This is in our usual notation of A(element) = log N(element)/N(H) + 12.00.)
As shown by Deliyannis et al.~(1998) the proposals to explain the Li dip by
mass loss (e.g. Schramm, Steigman \& Dearborn 1990) and by diffusion
(e.g. Michaud 1986) can be ruled out.  Wave-driven mixing (e.g.  Garc\'\i a
L\'opez \& Spruit 1991, Talon \& Charbonnel 2003) occurs closer to the surface
than does rotationally-induced mixing.  It is because these two mixing
mechanisms differ in their effectiveness with depth, that we were able to
distinguish between them with Li and Be measurements (Deliyannis et al.~1998).
Analogously, B offers a unique opportunity to examine whether this slow mixing
extends to deeper depths -- 10 - 30 \% by mass.

The inclusion of B in these light element studies allows us to 1) probe deeper
into stars, 2) search for B-depletions in individual stars, and 3) look for a
Be-B correlation, which could occur in the same temperature regime and be
similar in nature to the Li-Be correlation.  All of this will provide new
constraints for stellar models.  Our previous HST B study provided the
provocative hint that B is depleted in the severely Li- and Be-depleted star
$\zeta$ Her A (= HR 6212 = HD 150680).  Here we present B results for 13 newly
observed late F and early G dwarfs that are severely depleted in Be (and Li).
In addition, we have a sample of Li- and Be-normal stars to provide a
comparison group.  We are thus able to look at individual stars and to make a
statistical comparison between the normal stars and the Li-Be deficient ones.

\section{Observations}

\subsection{Star Selection}

As discussed above, it is significantly more difficult to deplete B than Be
because material must be circulated to greater depths (higher temperatures) to
destroy B atoms.  Therefore, in a search for B depletions, stars which are
severely depleted in Be are the most promising candidates.  We drew up our B
target list from stars with Be abundances determined in previous studies:
Deliyannis et al.~(1998) and Boesgaard et al.~(2001).  Relative to the
meteoritic value of A(Be) = 1.42, these stars are depleted in Be by factors of
10 - 50.  All the stars are also Li-depleted, but those depletions are so
extreme, that only upper limits can be determined for the Li abundances.  Our
sample of stars was also limited to F and G dwarfs and subgiants.

In addition we have observed B in five standard stars, i.e. those which would
not be expected to have depleted any Be or B.  Such stars would be those with
normal, undepleted Li.  To this group of standard stars we can add another
four Li-normal stars from the recent ``B Benchmarks'' paper by Boesgaard et
al.~(2004b) -- HR 740, HR 770, HR 4399, and HR 5927.

The positions of the stars in the HR diagram are shown in Figure 1 where the
standard stars are open squares and the program (Be-depleted) stars are filled
hexagons.  Shown for orientation is the zero-age main sequence for solar
metallicity.  The values for $T_{\rm eff}$ come from our determinations (see
section 3), while the values of log L/L$_{\odot}$ are from Hipparcos results.

\subsection{HST Observations of B}

We have made observations of the B I resonance line at 2497 \AA\ in 18 stars
with STIS aboard HST; they are listed in Table 1 which gives the dates and
exposure times.  Of these, four were bright enough to be observed with the
high-resolution E230H echelle grating with a 0$\arcsec$.2 $\times$
0$\arcsec$.09 aperture on the NUV-MAMA detector ($\lambda$/$\Delta$$\lambda$
$\sim$ 114,000) and the rest were observed with the medium-resolution E230M
echelle grating with a 0$\arcsec$.2 $\times$ 0$\arcsec$.06 aperture
($\lambda$/$\Delta$$\lambda$ = 30,000).  One orbit was dedicated to each star,
but two stars (HR 860 = HD 17948 and HR 7955 = HD 198084) were in the
``continuous viewing zone'' so longer exposures could be made of them.
Integration times range from 32 to 66 min and the signal-to-noise ratios per
pixel (S/N) range from 48 to 99 with a median of 70.  The STIS data were
reduced using the standard pipeline procedures.

\subsection{Li and Be Observations}

Although all of the standard stars had been previously observed for Li (by
Balachandran 1990 and Boesgaard \& Tripicco 1986), we obtained new
Keck/HIRES (Vogt et al.~1994) spectra at a spectral resolution of $\sim$48,000
of Li I resonance doublet is at 6708 \AA\ for four of them.  The fifth
standard, HR 8888 = HD 220242, had been observed with the UH 2.2-m telescope
and coud\'e spectrograph at a spectral resolution of $\sim$70,000 (Deliyannis
et al.~1998, Boesgaard et al.~2001).  We also observed the Li-Be depleted star
HR 107 (= HD 2454) anew at Keck/HIRES at high S/N to try to obtain a Li
detection, or at least a smaller upper limit for A(Li).

The same four standards were also observed at the Be II resonance lines (3130
and 3131 \AA) with Keck/HIRES.  The spectral resolution for these is
$\sim$48,000 and the effective pixel size is 0.022 \AA.  Results for Be in the
fifth standard star, HR 8888, are from a CFHT spectrum described in Boesgaard
et al.~(2001); it has a spectral resolution of 120,000 or a reciprocal
dispersion of 0.010 \AA\ pixel$^{-1}$.

The spectra were reduced using standard IRAF routines.  There were typically
13 flat-field files and 13 bias frames used to create the master flat and
master bias for each night.  Reference Th-Ar spectra were taken to calibrate
the wavelengths and determine the dispersion curve.

The log of the observations, including exposure times in sec and S/N ratios,
is given in Table 2.

\section{Abundances}

Abundances were determined for Li, Be, and B through the spectrum synthesis
mode of MOOG, 2002 version (Sneden 1973,
http://verdi.as.utexas.edu/MOOG.html).  This version contains the UV opacity
edges of the metals from Kurucz.

All of our stars had been observed previously for Be; it was those
observations that provided the selection criterion (severely-depleted Be) for
the stars observed for B.  The parameters of $T_{\rm eff}$, log $g$, [Fe/H],
and $\xi$ used for the model stellar atmospheres were adopted from those
papers: 12 stars from Boesgaard et al.~(2001) and two from Deliyannis et
al.~(1998).  The stellar parameters for the other four standard stars were
determined in the same manner as described in those papers.  The
microturbulent velocities, $\xi$, are derived from the empirical relation of
Edvardsson et al.~(1993).  The parameters for the stars are given in Table 3.

The grid of Kurucz (1993) model atmospheres was used for the benchmark models
to interpolate to get a model with the specific parameters for each star.

\subsection{Boron}

Examples of the spectrum synthesis fits are shown in Figure 2 where one of the
stars shown was observed at high spectral resolution and one at medium
resolution.  The best-fit B abundance is the solid line through the data
points.  The dashed lines show a factor of two more and a factor of two less
than the selected best value.  The B abundance in HR 761 is a factor of 2
lower than that in HR 860, while the Be abundance in HR 761 is a factor of 11
lower than in HR 860.  This shows that these stars have experienced different
degrees of mixing.

The abundances found by spectrum synthesis (in LTE) for B are given in column
9 in Table 4.  Corrections to LTE abundances for the effects of NLTE have been
calculated by Kiselman and Carlsson (1996).  These corrections have been
applied to our results and are listed in Table 4, column 10.  Note that the
range for the correction routines does not go above [Fe/H] = 0.00, so for our
eight stars with positive values for [Fe/H] we first inserted [Fe/H] = 0.0.
Trends with [Fe/H] from $-$0.20 to $-$0.10 to 0.0 for a few stars indicate
that the effect is negligible for our stars with [Fe/H] = +0.01 to +0.04 and
for HR 7955 at [Fe/H] = +0.15 the NLTE correction might be lower by 0.03 dex.
But for HR 2264 at [Fe/H] = +0.27, the NLTE-corrected value of A(B) might be
as low as 2.62 rather than 2.70 found at [Fe/H] = 0.0.  Such extrapolations
can be dangerous, but here they are small and affect only two stars, so it is
the extrapolated values for those two stars (HR 2264 and HR 7955) that appear
in Table 4.

\subsection{Lithium and Beryllium}

Examples of the Li and Be spectrum syntheses are shown in Figure 3 for the
standard star, HR 6394.  Unlike the other three newly observed standards,
these fits are excellent and the Li and Be abundances are well-determined.  We
found that HR 4803 was too hopelessly broadened by rotation to find a B
abundance, but both Li and Be could be determined (although with
larger-than-usual errors).  Although the spectra of HR 2264 and HR 6489 are
not as sharp-lined as those of HR 6394 and HR 8888, they could yield reliable
Li and Be abundances.  The Be abundances in Boesgaard et al.~(2001) were
redetermined from the original spectra with MOOG2002 which has the UV opacity
edges included; this procedure puts those spectra on the same standard as the
new ones for the standard stars.  The only significant changes were in A(Be)
for HR 2233 which increased from $-$0.08 to +0.22 and for HR 761 which
increased from $-$0.60 to $-$0.40.

\subsection{Error Estimates}

There are several potential sources of error in the abundances determinations.
Those associated with the errors in the stellar parameters are the most
straight-forward to estimate.  A typical uncertainty in the value of $T_{\rm
eff}$ of 80 K results in an uncertainty of 0.08 dex in A(B).  An uncertainty
in log $g$ of 0.2 produces an uncertainty in A(B) of 0.02 dex.  If the
metallicity is uncertain by 0.1 in [Fe/H], the resultant error in A(Be) is
0.03 dex.  These three parameters and their associated uncertainties are not
completely independent of each other nor are they closely interlinked; the
combined uncertainty would be between 0.09 (summed in quadrature) to 0.13
(addition of the errors).  Other sources of error include the S/N of the
observed spectrum, the details of the line list used in the fits, and what
could be called the ``goodness of fit'' estimate between the observed and
synthesized spectra.  The final estimates of the uncertainties are given in the
last column of Table 4.

In all cases the corrections for NLTE effects increase the B abundances.  They
range from +0.06 dex (for the metal-rich stars) to +0.22 dex(for the
metal-poorest stars in our sample).  We have applied these corrections as the
best available to us and use the NLTE results, but note that the corrections
can be sizeable.

The errors on Li and Be have been made in the papers from which those data are
drawn.  For the new observations of Li and Be in the standard stars here is a
synopsis of the effects of the uncertainties from the errors in the stellar
parameters for our range in temperatures, log g's and metallicities: for Be
$\Delta$$T$ = 80 K gives $\Delta$A(Be) = 0.01 dex; $\Delta$log $g$ = 0.1 gives
$\Delta$A(Be) = 0.04 dex (but increasing with decreasing temperature), and
$\Delta$[Fe/H] = 0.1 gives $\Delta$A(Be) = 0.04 dex.  For Li there is
virtually no effect on A(Li) from $\Delta$log $g$ = 0.1 or $\Delta$[Fe/H] =
0.1, but the change in temperature of 80 K gives $\Delta$A(Li) = 0.06 dex.

For Li the influence of NLTE effects is to decrease the derived LTE abundance.
For our sample of stars we have upper limits on A(Li) for all the Be-depleted
stars.  The NLTE corrections for the Li abundances in the five standard stars
are $-$0.10, $-$0.10, $-$0.09, $-$0.07 and $-$0.12 in order in Table 4.
According to Kiselman \& Carlsson (1996) and Garc\'\i a L\'opez, Severino \&
Gomez (1995) various effect of NLTE cancel each other for the Be II resonance
lines, so no corrections have been applied to A(Be).

\section{Results and Discussion}

\subsection{Boron Abundances and Depletions}

We can examine the degree of light element depletion by comparing the B and Be
abundances for our sample of stars.  The range in Be abundances (detections)
in our Be-depleted stars spans a factor of 11, while the range in B in these
same stars is nearly a factor of 5.  The difference in A(B) between the
highest and lowest value is 0.66 dex, well in excess of the individual errors,
typically 0.12 dex.  We can now look for a correlation between A(B) and A(Be)
with the expectation that the larger Be depletions will correlate with the
larger B depletions.  This is shown in Figure 4, where the left panel
corresponds to the LTE values of A(B) and the right panel the NLTE values.
The slopes for A(B) vs A(Be) in LTE and NLTE are virtually identical at 0.231
and 0.217, respectively, but there is a vertical offset of +0.16 dex in A(B)
due to the corrections for NLTE effects.  Although the points with upper
limits on A(Be) are plotted (as small open circles with left leading arrows),
they are not included in the least-squares fit.  The relationship is between
depleted, but detected, Be and B abundances.

This indicates the discovery of a correlation between the Be and B abundances,
similar to the correlation between Li and Be.  The slope between Be and B of
0.22 is shallower than that between Li and Be which is 0.38 (from Figure 9 of
Boesgaard et al.~2004a).  This can provide clues about how the mixing changes
with depth.  The B atoms have to be circulated to greater depths to be
destroyed and therefore B is the less susceptible to destruction.

To demonstrate the reality of the B depletion, we show the spectrum synthesis
fit for the star with the largest Be deficiency, HR 107.  Figure 5 shows the B
I region and the syntheses for HR 107 in the upper panel.  For comparison a
star of similar metallicity was chosen from the sample of Boesgaard et
al.~(2004b), HD 11592.  The synthesis fit has 5 times less B in HR 107 than in
HD 11592 and the difference in the line depth of the B feature is obvious.

Tomkin et al.~(1989) describe HR 107 as ``an F-type mild barium dwarf star.''
It has enhanced s-process products as Y, Zr, Ba, and Ce relative to HR 366, a
star of similar temperature, gravity, and metallicity.  The Ba star phenomenon
apparently arises from mass transfer from an AGB star (which is now a white
dwarf) to a less massive companion (e.g. McClure, Fletcher \& Nemec 1980).
Although no white dwarf companion has been detected near HR 107, its
compositional peculiarities are symptomatic of Ba stars.  The mass transfer
process, which pollutes the atmosphere with material processed in the interior
of the AGB star, could affect the light element abundances.  Its Li and Be are
undetectable and it is deficient in B.

\subsection{Correlation of Beryllium and Boron}

Figure 4 shows the interesting correlation of Be and B similar to the Li-Be
correlation first discovered by Deliyannis et al.~(1998) and advanced by
additional observations of Boesgaard et al.~(2001) and Boesgaard et
al.~(2004c).  To examine this further we decided to 1) restrict the
temperature range to 5900 - 6650 K (as was done for the Li-Be correlation), 2)
include the Li-normal stars in Boesgaard et al.~(2004b) and 3) normalize the
abundances to solar metallicity.

It was found in Boesgaard et al.~(2004b) that both A(Be) and A(B) increase
with [Fe/H] in the Galactic disk.  This result is based on HST/STIS
observations of B in 20 stars with ``undepleted'' Be abundances from spectra
taken at KeckI with HIRES or at CFHT with the Gecko spectrometer.  The stars
were chosen because they fell on the upper envelope in a plot of A(Be) vs
[Fe/H] that covered a range of [Fe/H] from $-$1.0 to +0.2.  The light element
abundances in these stars could be well represented by these relationships for
stars in the Galactic disk:

A(Be) = (0.382 $\pm$0.135)[Fe/H] + 1.218 $\pm$0.037

A(B)$_{NLTE}$ = (0.402 $\pm$0.117)[Fe/H] + 2.371 $\pm$0.032

In order to estimate the true depletion of Be and B in our stars we need to
make an adjustment for the ``initial'' Be and B, which appears to depend on
metallicity (or age).  A star that has fewer metals, e.g. [Fe/H] = $-$0.30,
would start with somewhat lower Be and B than a star of solar metallicity.
According to the above relationships, a star at [Fe/H] = 0.0 would have A(Be)
= 1.22 and A(B)$_{NLTE}$ = 2.37, while the star with [Fe/H] = $-$0.30 would
start with A(Be) = 1.10 and A(B)$_{NLTE}$ = 2.25.  So the depletion caused by
internal slow mixing should be taken relative to that starting value.  So we
have ``normalized'' the Be and B abundances to solar metallicity.

Figure 6 shows the correlation of Be and NLTE-B with the normalized values.
As expected, the slope is a bit shallower because some component of the lower
abundance for Be and B is not real depletion, but caused by a lower initial Be
and B.  The relationship from this group of stars (taking into account the
errors in both coordinate) is:

A(B)$_{NLTE}$ = (0.180 $\pm$0.063)A(Be) + 2.219 $\pm$0.060

We note that although the slopes are fairly shallow the errors are small and
the result is significant at the 3 $\sigma$ level.  The linear correlation
coefficient indicates a correlation significant at the 96.5\% level for the
above relationships.  

Another approach to elucidate the B-Be correlation is via statistics.  We can
compare the abundances in two groups.  One group is composed of the standard
stars, or ``Li-normal stars'' which have small or no depletion in all three
elements from this paper and from Boesgaard et al.~(2004a).  The other group
contains the six stars with large Be depletions of over an order of magnitude.
These stars are listed in the two groups in Table 5 along with their mean
values of the normalized Be and B abundances.  Here the errors are the
standard deviation of the mean.  The groups were binned by the Be data and the
difference in the mean B abundances is 0.22 dex, well in excess of the
standard deviations by at least 4 $\sigma$.  One of our goals was to ascertain
the reality of B depletions by statistical means, but this turns out to be a
statistical confirmation of the Be-B correlation.  We have not included HR
7697 and HR 8792 in Table 5 because of their upper limit measurements on
A(Be), but they do have B detections.  Including them in the normalized B list
at 2.19 and 1.96 gives a mean for the eight Be-deficient stars of 2.16
$\pm$0.04.  This is difference between the Li-normal group and the Be-depleted
group of 0.25 dex, a factor of 1.8.

\section{Summary and Conclusions}

We have conducted a study of B in stars with large Be depletions to ascertain
whether the mixing in such stars goes down to great enough depths to destroy B
also.  Such information is needed to provide new constraints on stellar models
because standard models predict no depletion at all and the emerging new set
of more sophisticated models require additional observational information for
validation.  We have examined the evidence for a possible correlation between
Be and B analogous to the correlated depletions of Li and Be.

We selected 13 F and early G dwarfs with moderate to severe Be depletions and
five stars with undepleted Li for comparison. Since Li is the most sensitive
to destruction, stars with normal Li are expected to have no Be or B
depletion.  One orbit of HST was used for each star; the median S/N at the B I
resonance line was 70.  Complementary ground-based observations of Li and Be
were made for the standard stars, while results on Li and Be from previous
papers by Deliyannis et al. (1998) and Boesgaard et al. (2001) were used for
the program stars.  Abundances of all three elements were determined by the
spectrum synthesis method of MOOG2002.

The program stars show a range in Be depletions of a factor of 11 and B
depletions range up to nearly a factor of 5.  The abundances of B and Be are
correlated, with a slope of $\sim$0.22.  We have normalized the Be and B
abundances to take into account that the initial abundances may have been
lower for the low metallicity stars, using the relationships discovered by
Boesgaard et al.~(2002b) for Galactic disk stars.  This results in a somewhat
shallower slope, 0.18 $\pm$0.06, but a clear trend.

We also approach this issue by binning the stars into two groups and
determining mean Be and B abundances.  The groups are seven Li-normal stars
(A(Li) $\geq$3.0 from this paper and from Boesgaard et al.~(2004b) and six
stars with A(Be) $<$0.3.  The B abundances in the two groups are A(B)$_{NLTE}$
(normalized) are 2.41 $\pm$0.05 and 2.19 $\pm$0.04.  As expected the B
deficiencies are smaller than the Be deficiences (which in turn are smaller
than the Li deficiencies), but the existence of B depletions is confirmed.

The star with the most severe Be depletion, HR 107 at A(Be) $<$ $-$1.00, has
the largest B depletion, a detection of B at A(B)$_{LTE}$ = 1.50 and
A(B)$_{NLTE}$ = 1.72; these values become A(Be) $<$$-$0.86 and A(B)$_{NLTE}$ =
1.87 when normalized to solar [Fe/H].  However, HR 107 is known to be a
main-sequence Ba star with enhanced s-process elements.  The light element
abundances would have been altered in the process of the mass transfer that
resulted in the characteristic Ba star symptoms.  The normalized Be depletion
is more than 2 orders of magnitude and the normalized B depletion is more than
a factor of 2.

The correlation of Be and B should be included as a constraint on models of F
and G dwarfs -- in particular those including the effects of stellar rotation.
Rotationally-induced mixing has been successful in reproducing Li and Be
abundances and that correlation for stars with $T_{\rm eff}$ from 5900 - 6300
K as seen in Figure 12 in Boesgaard et al.~(2004c).

\acknowledgments We are pleased to acknowledge the industrious and careful
work of John Lakatos on the data reduction for this project.  This work has
been supported by grant HST-GO-09048 to CPD and to AMB and by NSF grant AST
00-97945 to AMB.

\clearpage

\begin{figure}
\plotone{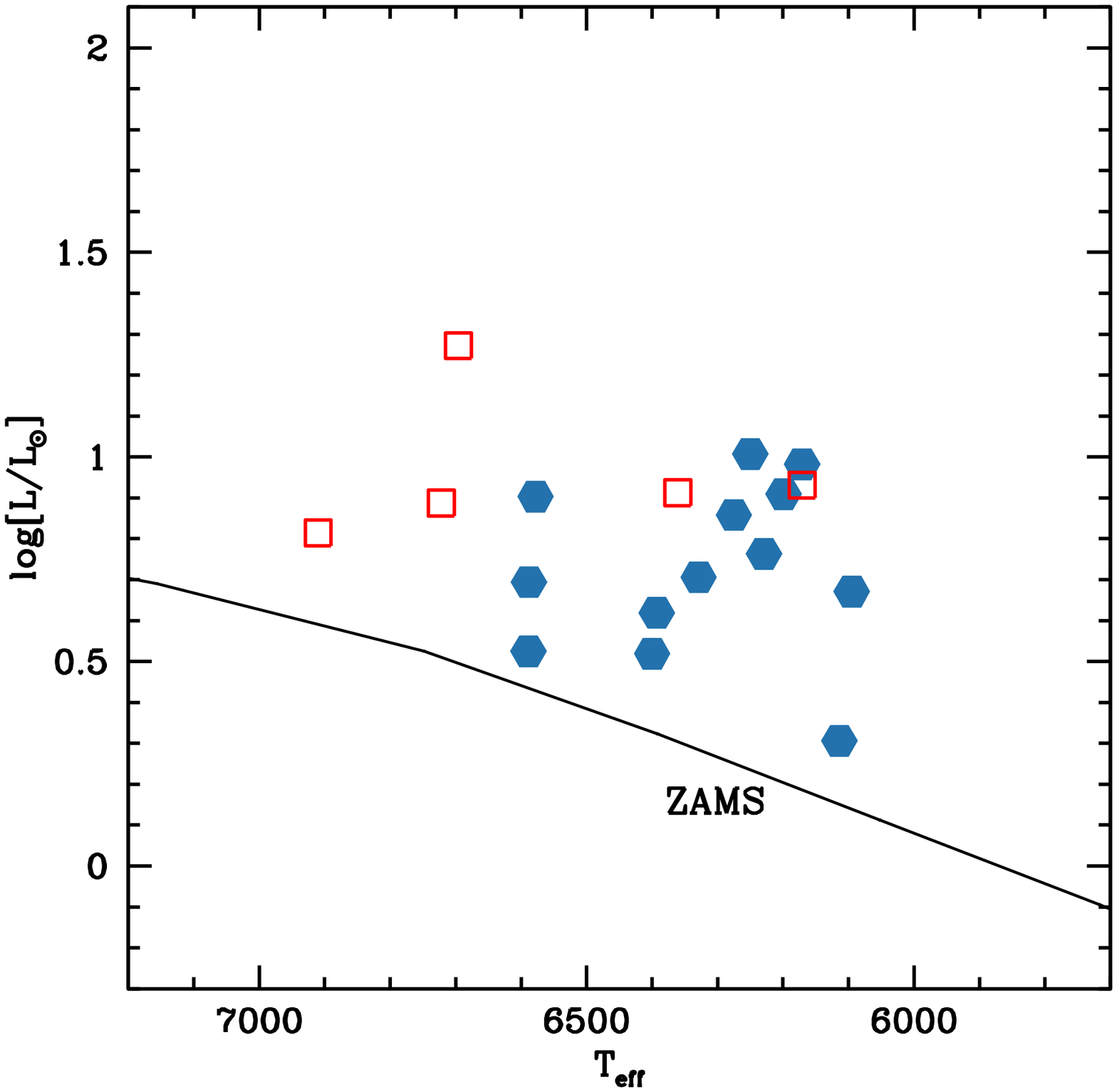}
\caption{HR diagram showing the positions of our stars.  The filled hexagons
are the Be-depleted stars while the open squares are the standard stars with
normal Li and Be abundances.  The ZAMS track is from the Yale isochrones for
solar composition.}
\end{figure}

\begin{figure}
\plotone{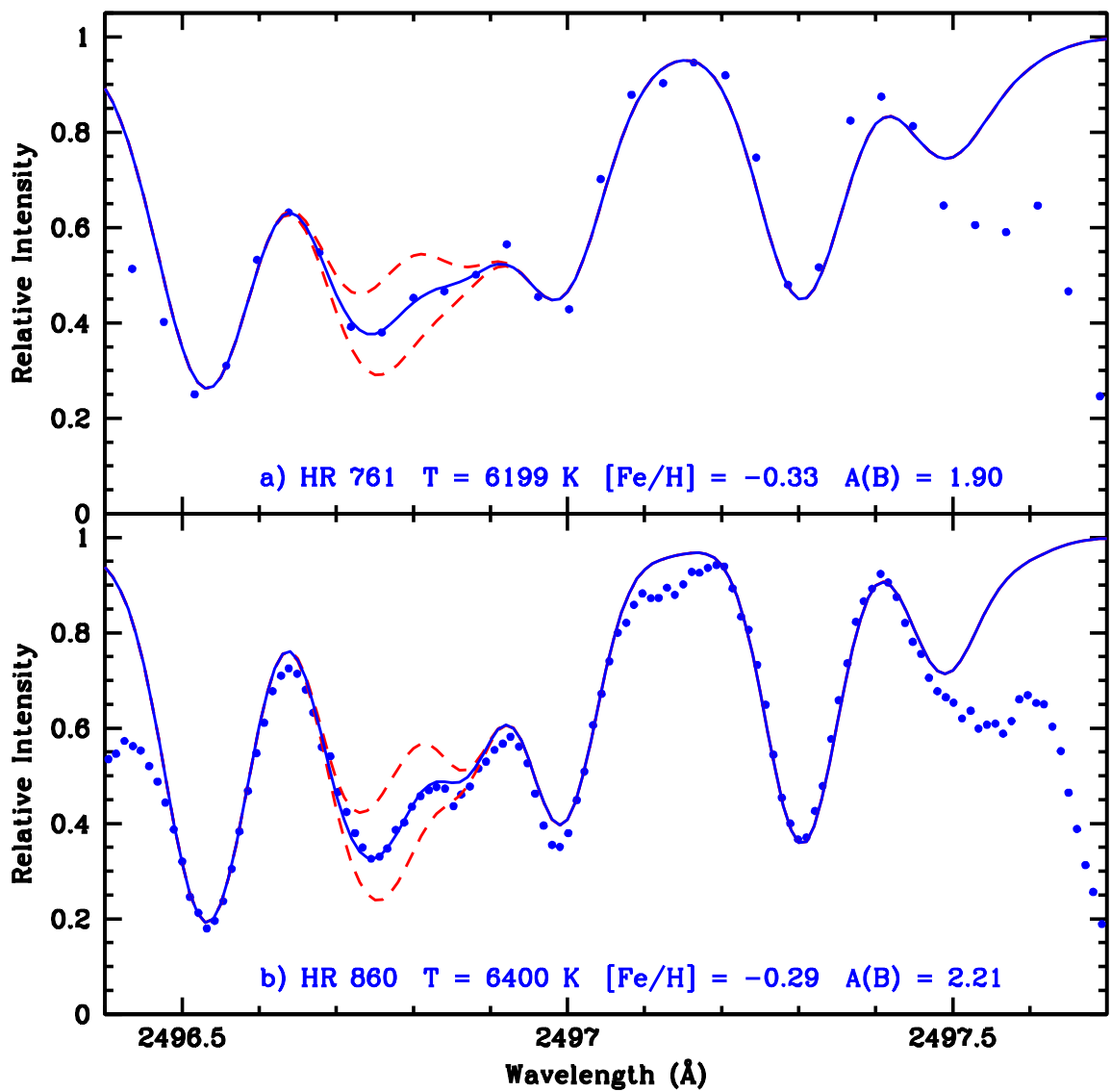}
\caption{Representative B I region spectra.  The upper panel showing HR 761 is
a medium resolution spectrum (30,000), while the lower panel showing HR 860 is
a high resolution spectrum (114,000).  The dots are the observed spectral data
points, the solid line is the best synthetic fit.  The dashed lines represent
a factor of two above and below the best B abundance.  These two stars have
similar metallicities, but their temperatures and log g values are different
as can be seen in their respective positions in Figure 1; HR 761 appears to
have evolved off the ZAMS from a prior locale in the Li-Be dip region.  The B
abundances differ by a factor of 2.}
\end{figure}

\begin{figure}
\plotone{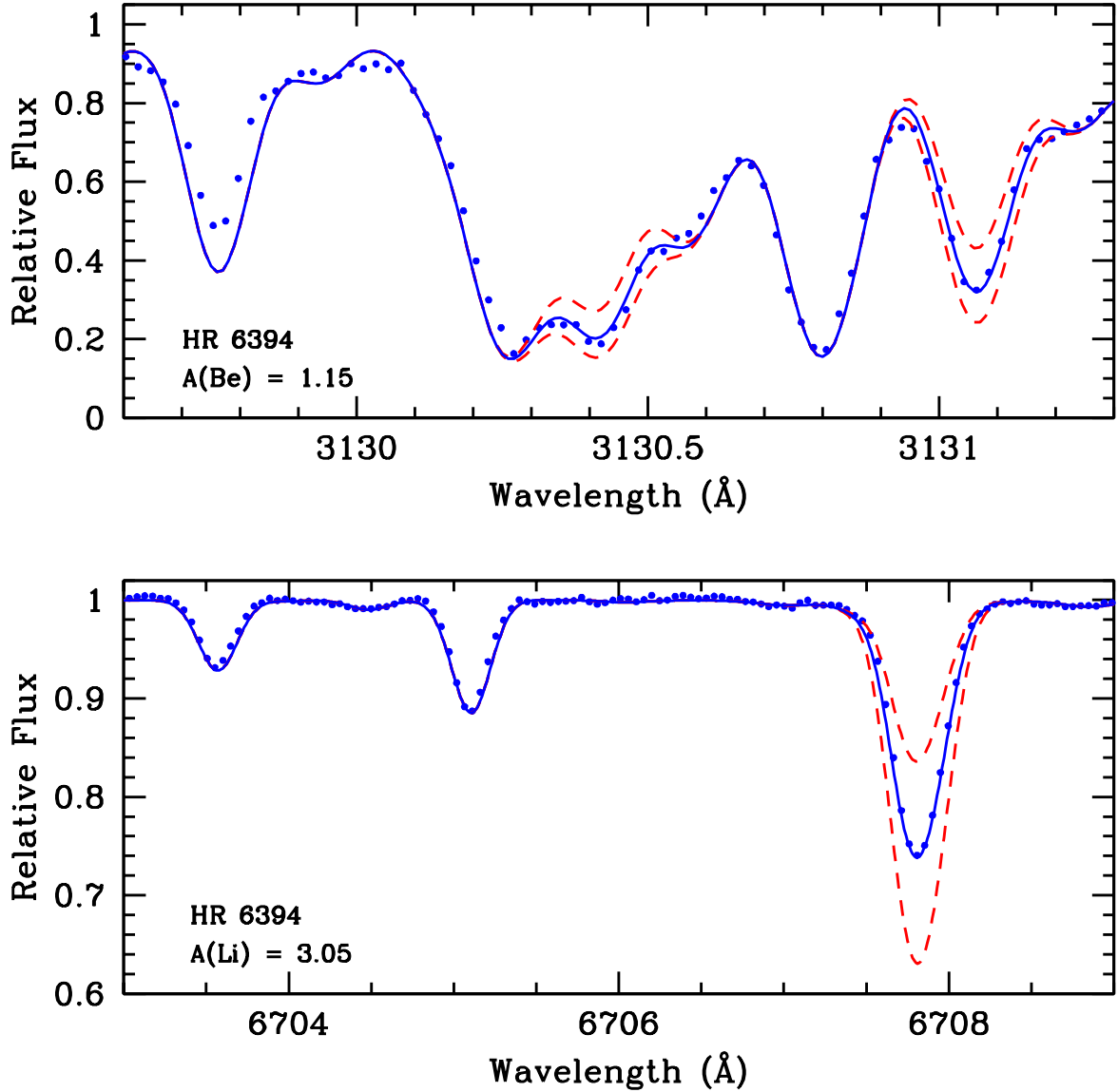}
\caption{Newly obtained Be and Li spectra for one of our standard stars.  The
Be spectrum has S/N per pixel of 37 while the Li spectrum has 530.  The
observations are the dots and the solid line is the best spectrum synthesis
fit.  The dashed lines are a factor of 2 above and below the best fit.}
\end{figure}

\begin{figure}
\plottwo{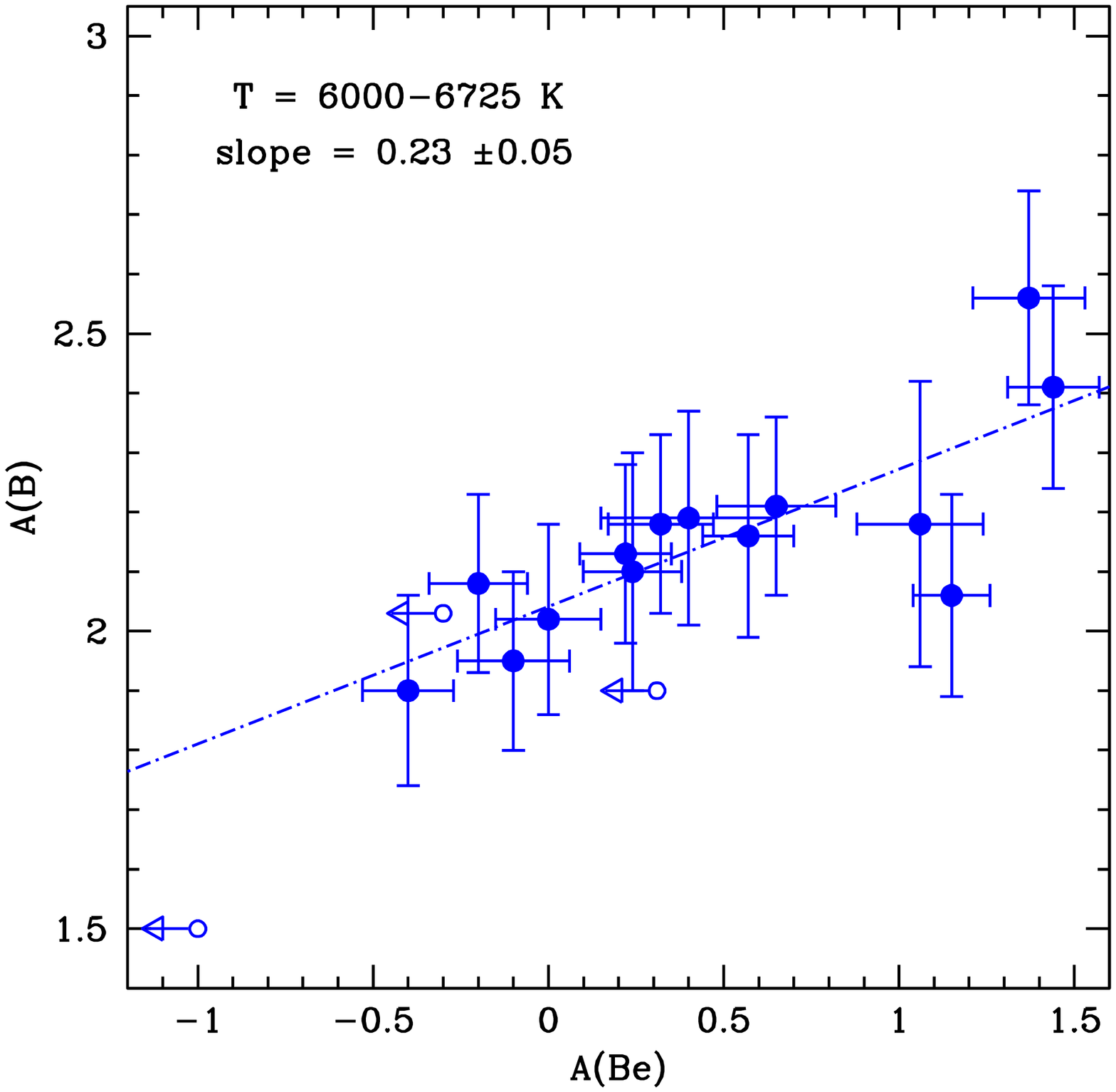}{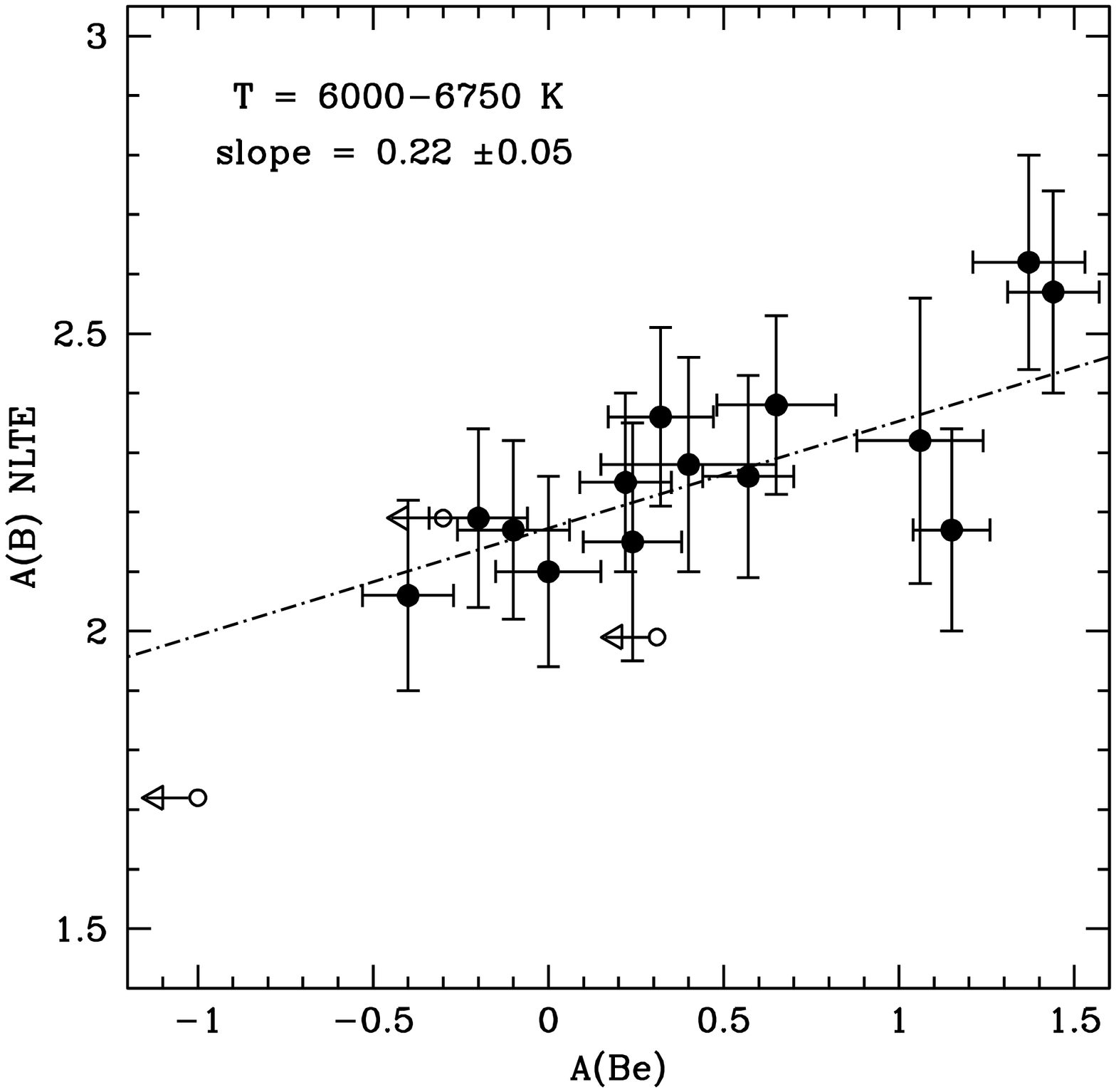}
\caption{The newly-discovered correlation between the abundances of Be and B,
the left panel is from the LTE B results while the the right panel is
corrected for NLTE effects.  Most of our stars have moderate to severe Be
depletions and show accompanying mild to moderate B depletions.  The slope of
this relationship is 0.23 $\pm$0.05 (LTE B) to 0.22 $\pm$0.05 (NLTE B).  Three
of these stars have B detections and determinations, but have only upper
limits on the Be abundances (HR 107, HR 7697, and HR 8792); these are shown as
small open circles with arrows pointing to lower Be values.  These three stars
were not used in the calculation of the slope.}
\end{figure}

\begin{figure}
\plotone{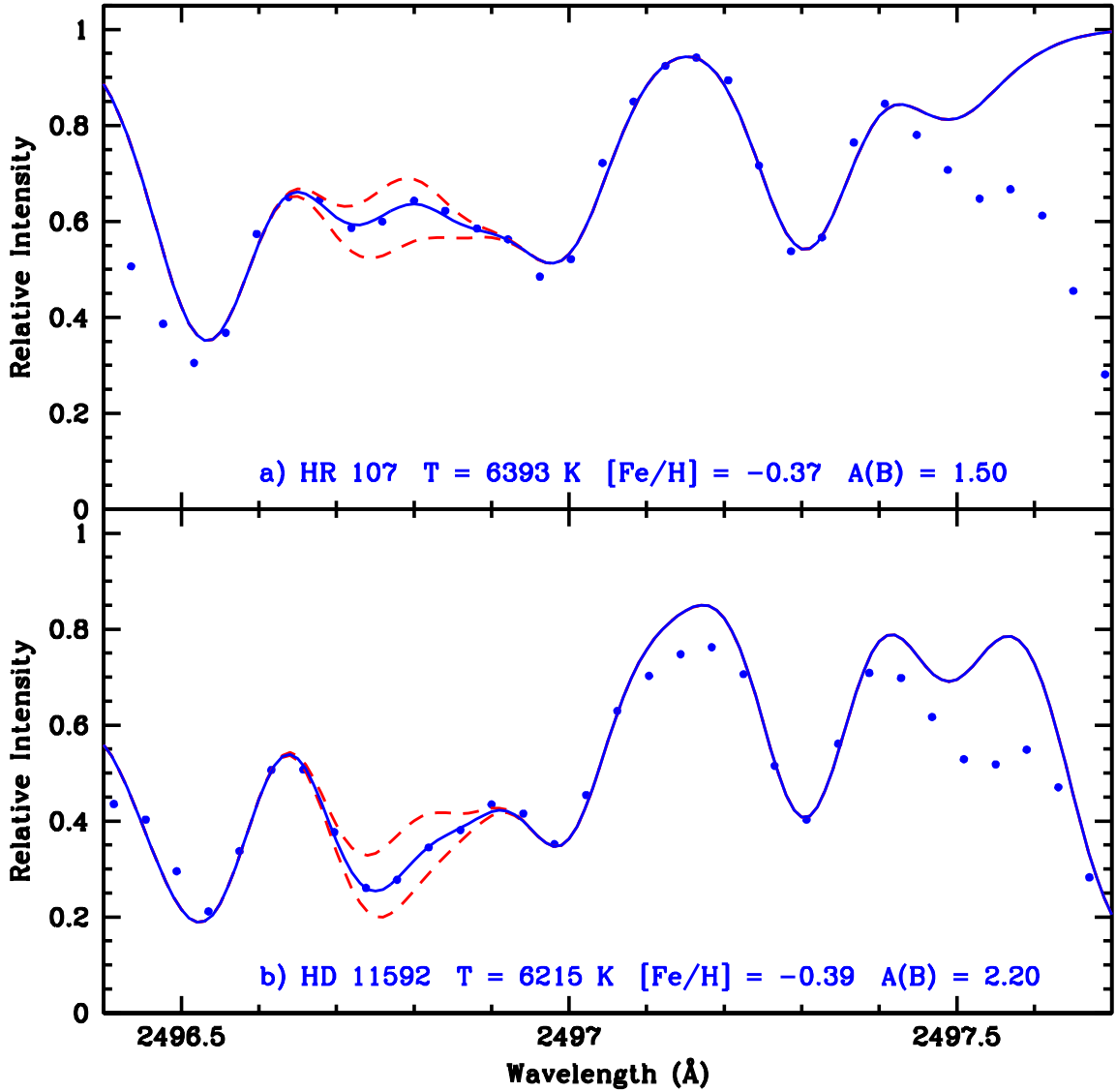}
\caption{The B I region spectrum of the most severely Be-depleted star in our
sample, HR 107.  The spectrum synthesis fit for B in HR 107 is compared here
with that of a star of similar metallicity and temperature, HD 11592, from
Boesgaard et al. (2004).  The B depletion is obvious in the shape of the
observed spectrum as well as in the best-fit synthesis.}
\end{figure}

\begin{figure}
\plotone{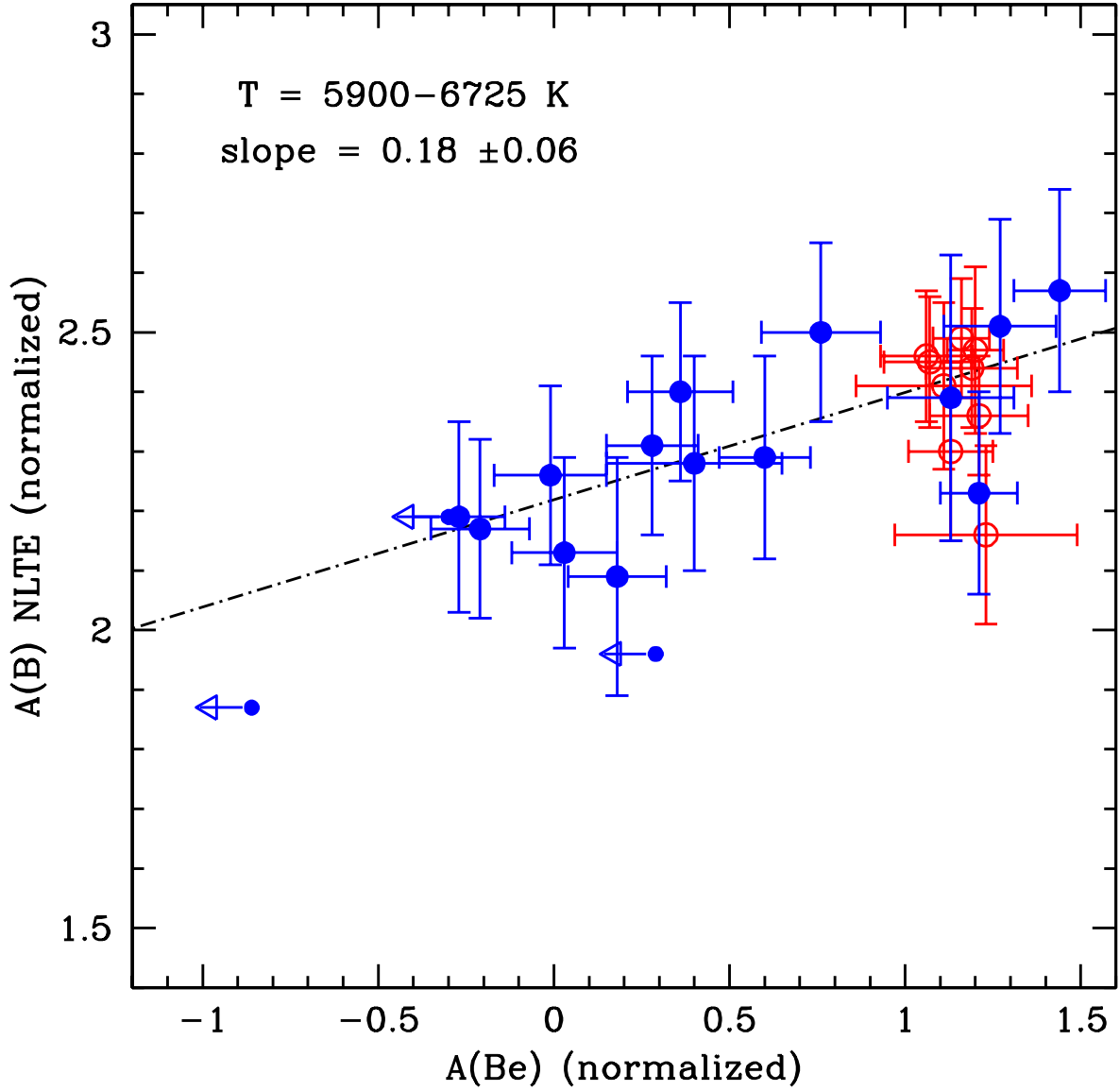}
\caption{Normalized Be and B abundances (see text $\S$4.2).  The solid symbols
are from this paper and the open circles are from Boesgaard et al.~(2004b).
The slope includes the errors in both coordinates}
\end{figure}

\clearpage

\topmargin=-1.25in
\begin{deluxetable}{rrccrcc} 
\tablenum{1}
\tabletypesize{\normalsize}
\tablecolumns{7} 
\tablewidth{0pc} 
\tablecaption{Observations of Boron \label{boron-tbl}} 
\tablehead{ 
& \colhead{Star ID} & & & \colhead{Spectral} & \colhead{exp. time} &  \\
\colhead{HR} & \colhead{HD} & $V$ & \colhead{Date Observed} & 
\colhead{Resolution} & \colhead{(sec.)} & \colhead{S/N} }
\startdata 
107  &   2454  & 6.06 & 2001 Oct 30 & 30,000  & 1929 &  88 \\
638  &  13456  & 6.01 & 2002 Feb 10 & 30,000  & 1929 &  97 \\
761  &  16220  & 6.25 & 2002 Feb 06 & 30,000  & 1975 &  72 \\
860  &  17948  & 5.59 & 2002 Aug 24 & 114,000 & 3960 &  81 \\
988  &  20395  & 6.14 & 2002 Jan 25 & 30,000  & 1929 &  90 \\
2233 &  43318  & 5.65 & 2002 Feb 16 & 30,000  & 1929 &  85 \\
2264 &  43905  & 5.36 & 2001 Nov 16 & 114,000 & 2144 &  52 \\
4803 & 109799  & 5.45 & 2002 Jul 19 & 114,000 & 1956 &  48 \\
6394 & 155646  & 6.65 & 2002 Sep 11 & 30,000  & 1927 &  51 \\
6489 & 157856  & 6.44 & 2001 Sep 20 & 30,000  & 1927 &  68 \\
7454 & 184985  & 5.47 & 2001 Sep 15 & 30,000  & 1929 &  81 \\
7658 & 190009  & 6.45 & 2001 Sep 18 & 30,000  & 1934 &  57 \\
7697 & 191195  & 5.85 & 2001 Oct 12 & 30,000  & 2141 &  99 \\
7955 & 198084  & 4.51 & 2001 Nov 28 & 114,000 & 3960 &  64 \\
8077 & 200790  & 5.94 & 2001 Oct 17 & 30,000  & 1929 &  61 \\
8250 & 205420  & 6.47 & 2001 Nov 07 & 30,000  & 1934 &  53 \\
8792 & 218261  & 6.30 & 2001 Nov 03 & 30,000  & 1927 &  51 \\
8888 & 220242  & 6.62 & 2001 Nov 29 & 30,000  & 1951 &  75 \\
\enddata 
\end{deluxetable} 


\clearpage

\topmargin=-1.25in
\begin{deluxetable}{rrcccccc} 
\tablenum{2}
\tabletypesize{\normalsize}
\tablecolumns{8} 
\tablewidth{0pc} 
\tablecaption{New Keck/HIRES Observations of Lithium and Beryllium}
\tablehead{ 
 \colhead{Star ID} & &  \colhead{Lithium}   & & &
\colhead{Beryllium} & &  \\
\colhead{HR} & \colhead{HD} & \colhead{Date} & \colhead{exp. (sec.)} & 
\colhead{S/N} & \colhead{Date} & \colhead{exp. (sec.)} & 
\colhead{S/N}
 }
\startdata 
107  & 2454   & 2002 Jan 11 & 840  & 790 & \nodata & \nodata & \nodata \\
2264 & 43905  & 2002 Jan 11 & 120  & 435 & 1995 Mar 14 & 300  & 90  \\
4803 & 109799 & 2002 Jan 11 & 120  & 575 & 2001 Feb 01 & 600  & 55 \\
6394 & 155646 & 2002 Feb 11 & 240  & 530 & 2001 Feb 01 & 600  & 37 \\
6489 & 157856 & 2002 Feb 11 & 240  & 505 & 2001 Feb 01 & 600  & 41 \\
\enddata 
\end{deluxetable} 


\clearpage

\topmargin=-1.25in
\begin{deluxetable}{rrccccc}
\tablenum{3}
\tabletypesize{\normalsize}
\tablecolumns{7}
\tablewidth{0pc}
\tablecaption{Stellar Parameters}
\tablehead{
\colhead{HR} & \colhead{HD} & \colhead{log L/L$_{\odot}$} &
\colhead{$T_{\rm eff}$}  & \colhead{log $g$} & \colhead{[Fe/H]}  & 
\colhead{$\xi$(km s$^{-1})$} 
}
\startdata
\cutinhead{Be-Depleted Stars \phm{blahhhhhhhhhhhhhhhhhhhh}} 
107	& 2454	 & 0.619 & 6393	& 4.35	& $-$0.37	& 1.8 \\
638     & 13456	 & 0.903 & 6578	& 4.12	& $-$0.23	& 2.2 \\
761	& 16220	 & 0.909 & 6199	& 3.95	& $-$0.33	& 2.1 \\
860	& 17948	 & 0.519 & 6400	& 4.16	& $-$0.29	& 2.0 \\
988	& 20395	 & 0.525 & 6589	& 4.29	& $-$0.09	& 2.0 \\
2233	& 43318	 & 0.763 & 6229	& 4.18	& $-$0.16	& 1.7 \\
7454	& 184985 & 0.706 & 6329	& 4.18	& +0.04		& 1.9 \\
7658	& 190009 & 0.858 & 6275	& 4.00	& +0.01		& 2.1 \\
7697	& 191195 & 0.694 & 6588	& 4.24	& +0.01		& 2.1 \\
7955	& 198084 & 0.982 & 6171	& 4.13	& +0.15		& 1.9 \\
8077	& 200790 & 0.671 & 6095	& 4.08	& $-$0.07	& 1.9 \\
8250	& 205420 & 1.007 & 6250	& 4.03	& $-$0.07	& 2.1 \\
8792	& 218261 & 0.306 & 6114	& 4.20	& +0.06		& 1.6 \\
\cutinhead{Standard Stars \phm{blahhhhhhhhhhhhhhhhhhhhhh }}
2264	& 43905	 & 1.272 & 6696	& 3.65	& +0.27		& 2.9 \\
4803	& 109799 & 0.814 & 6910	& 4.20	& +0.01		& 2.7 \\
6394	& 155646 & 0.931 & 6171	& 3.91	& $-$0.14	& 2.2 \\
6489	& 157856 & 0.913 & 6361	& 4.00	& $-$0.17	& 2.2 \\
8888    & 220242 & 0.887 & 6722	& 4.00	& +0.01		& 2.5 \\

\enddata

\end{deluxetable}

\clearpage

\topmargin=-1.25in
\begin{deluxetable}{rrcccrcccccrc}
\tablenum{4}
\tabletypesize{\scriptsize}
\tablecolumns{13}
\tablewidth{0pc}
\tablecaption{Abundance Results}
\tablehead{
\colhead{HR} & \colhead{HD} & \colhead{$T_{\rm eff}$}  
& \colhead{A(Li)}   & \colhead{ref\tablenotemark{1}}
& \colhead{A(Be)}  & \colhead{$\sigma$}  & \colhead{ref\tablenotemark{1}}
& \colhead{A(B)$_{LTE}$}  & \colhead{A(B)$_{NLTE}$}  & \colhead{$\sigma$} 
&\colhead{A(Be)-norm.}  &\colhead{A(B)$_{NLTE}$-norm.}
}
\startdata
\cutinhead{Be-Depleted Stars \phm{blahhhhhhhhhhhhhhhhhhhhhhhhhhhhhhhhhhhhhhh}} 
107  & 2454   & 6393 & $<$0.81 & new & $<-$1.00 & 0.16 & a   & 1.50 & 1.72 &
0.15 & $<-$0.86 & 1.87 \\
638  & 13456  & 6578 & $<$1.13 & a   & $-$0.10 & 0.16 & a   & 1.95 & 2.17 &
0.15 & $-$0.01 & 2.26 \\
761  & 16220  & 6199 & $<$0.67 & a   & $-$0.40 & 0.13 & a   & 1.90 & 2.06 &
0.16 & $-$0.27 & 2.19 \\
860  & 17948  & 6400 & $<$0.82 & a   & 0.65    & 0.17 & a   & 2.21 & 2.38 &
0.15 & 0.76    & 2.50 \\
988  & 20395  & 6589 & $<$1.36 & a   & 0.32    & 0.15 & a   & 2.18 & 2.36 &
0.15 & 0.36    & 2.40 \\
2233 & 43318  & 6229 & $<$0.86 & a   & 0.22    & 0.13 & a   & 2.06 & 2.25 &
0.15 & 0.28 & 2.31 \\
7454 & 184985 & 6329 & $<$1.17 & a   & $-$0.20 & 0.14 & a   & 2.08 & 2.19 &
0.15 & $-$0.21 & 2.17 \\
7658 & 190009 & 6275 & $<$0.96 & b   & 0.40    & 0.25 & b   & 2.19 & 2.28 &
0.18 & 0.40    & 2.28 \\
7697 & 191195 & 6588 & $<$0.83 & a   & $<-$0.30 & 0.15 & a   & 2.03 & 2.19 &
0.19 & $<-$0.30 & 2.19 \\
7955 & 198084 & 6171 & $<$0.56 & a   & 0.24    & 0.14 & a   & 2.10 & 2.15 &
0.20 & 0.18    & 2.09 \\
8077 & 200790 & 6095 & $<$0.58 & a   & 0.00    & 0.15 & a   & 2.02 & 2.10 &
0.16 & 0.03    & 2.13 \\
8250 & 205420 & 6250 & $<$0.72 & a   & 0.57    & 0.13 & a   & 2.16 & 2.26 &
0.17 & 0.60    & 2.29 \\
8792 & 218261 & 6114 & $<$0.96 & b   & $<$0.31 & 0.25 & b   & 1.90 & 1.99 &
0.18 & $<$0.29 & 1.96 \\
\cutinhead{Standard Stars \phm{blahhhhhhhhhhhhhhhhhhhhhhhhhhhhhhhhhhhhhhhhh}}
2264 & 43905  & 6696 & 3.18    & new & 1.37    & 0.16 & new & 2.56 & 2.62 &
0.18 & 1.27    & 2.51 \\
4803 & 109799 & 6910 & 3.15    & new & 1.09    & 0.27 & new & \nodata &
\nodata & \nodata & 1.09    & \nodata \\ 
6394 & 155646 & 6171 & 3.05    & new & 1.15    & 0.11 & new & 2.06 & 2.17 &
0.17 & 1.21    & 2.23 \\
6489 & 157856 & 6361 & 2.88    & new & 1.06    & 0.18 & new & 2.18 & 2.32 &
0.24 & 1.13    & 2.39 \\
8888 & 220242 & 6722 & 3.30    & a   & 1.44    & 0.13 & a   & 2.41 & 2.57 &
0.17 & 1.44    & 2.57 \\

\enddata 
\tablenotetext{1}{new - see Table 2; a = Boesgaard et al.(2001) as
redetermined from MOOG2002; b =
Deliyannis et al. (1998)}
\end{deluxetable}

\clearpage

\topmargin=-1.25in
\begin{deluxetable}{rrcccccc}
\tablenum{5}
\tabletypesize{\normalsize}
\tablecolumns{8}
\tablewidth{0pc}
\tablecaption{Abundances in Li-Normal Stars and Be-Depleted Stars}
\tablehead{
\colhead{HR} & \colhead{HD} & \colhead{$T_{\rm eff}$}  & \colhead{[Fe/H]} &
A(Li) & A(Be)-norm. & \colhead{A(B)$_{NLTE}$-norm.} 
& \colhead{ref\tablenotemark{1}}
}
\startdata
\cutinhead{Li-Normal Standard Stars\phm{blahhhhhhhhhhhhhhhhhhhhhhhhhhhhhhhh }} 
2264    & 43905  & 6696 &   +0.27 & 3.18 & 1.27 & 2.51 & a \\
6394    & 155646 & 6171 & $-$0.14 & 3.05 & 1.21 & 2.23 & a \\
8888    & 220242 & 6722 &   +0.01 & 3.30 & 1.44 & 2.57 & a \\
740     & 15798  & 6347 & $-$0.22 & 3.14 & 1.19 & 2.44 & b \\
770	& 16399  & 6481 & $-$0.08 & 3.13 & 1.21 & 2.36 & b \\
4399	& 99028  & 6679 & +0.11   & 3.30 & 1.03 & 2.32 & b \\
5927	& 142640 & 6387 & +0.05   & 3.35 & 1.20 & 2.47 & b \\
mean    &        &      &         &      & 1.22 $\pm$0.05 & 2.41 $\pm$0.05 \\
\cutinhead{Be-Depleted Stars\phm{blahhhhhhhhhhhhhhhhhhhhhhhhhhhhhhhhhhhhhhh }} 
638     & 13456  & 6578 & $-$0.23 & $<$1.13 & $-$0.01 & 2.26 & a \\
761     & 16220  & 6199 & $-$0.33 & $<$0.67 & $-$0.27 & 2.19 & a \\
2233    & 43318  & 6229 & $-$0.16 & $<$0.86 & +0.28   & 2.31 & a \\
7454    & 184985 & 6329 & +0.04   & $<$1.17 & $-$0.21 & 2.17 & a \\
7955    & 198084 & 6171 & +0.15   & $<$0.56 & +0.18   & 2.09 & a \\
8077    & 200790 & 6095 & $-$0.07 & $<$0.58 & +0.03   & 2.13 & a \\
mean    &        &      &         &        & 0.00 $\pm$0.21 & 2.19 $\pm$0.04 \\
\enddata
\tablenotetext{1}{a = This paper; b = Boesgaard et al.(2004b)}
\end{deluxetable}

\clearpage

\end{document}